\documentclass[aps,prl,twocolumn,groupedaddress]{revtex4}
\usepackage{graphicx,latexsym}

\begin{document}

\title{The strengthening of reentrant pinning by collective interactions in the peak effect}

\author{J.\ Lefebvre, M.\ Hilke, and Z.\ Altounian}

\affiliation{ Department of Physics, McGill University, Montr\'eal, Canada
H3A 2T8.}

\begin{abstract}

Since it was first observed about 40 years ago \cite{Pippard}, the
peak effect has been the subject of numerous research mainly impelled by the
desire to determine its exact mechanisms. Despite these efforts, a consensus on this question has yet to be reached. Experimentally, the peak effect indicates a
transition from a depinned vortex phase to a reentrant pinning phase at high
magnetic field. To study the effects of intrinsic pinning on the peak effect, we consider Fe$_{x}$Ni$_{1-x}$Zr$_{2}$ superconducting metallic glasses in which the vortex pinning
force varies depending on the Fe content and in which a huge peak effect is seen as a function of magnetic field. The results are mapped out as a phase diagram in which it is readily seen that the peak effect becomes broader with decreasing pinning force. Typically, pinning can be understood by increased pinning centers, but here, we show that reentrant pinning is due to the strengthening of interactions (while decreasing pinning strength). Our results demonstrate the strengthening of the peak effect by collective effects.
\end{abstract}

\maketitle

Vortices in type II superconductors form a correlated system of interacting
particles which can be studied as a function of particle density or driving
force by simply tuning the external magnetic field or driving current. While
elastic vortex-vortex interactions tend to order the system, vortex-pin
interactions result in disorder. An ever intriguing phenomena resulting from
this competition between elastic and plastic interactions is the peak effect
(PE): an anomalous peak in the critical current $J_{c}$ (or dip in resistance)
appearing with increasing temperature or magnetic field just below the
transition to the normal state in some conventional superconductors
\cite{BerlincourtPRL6, RosenblumRMP36, Pippard, KesPRB28, PaltielPRL85,
HilkePRL91}, and at lower field below B$_{c2}$ in high T$_{c}$ superconductors \cite{KwokPRL73}. In type II superconductors, vortices will depin under
the action of a driving force larger than the critical force. As a result of
vortex motion, a dissipative voltage proportional to $\mathbf{E=\bar{v}\times
B}$ where $\mathbf{\bar{v}}$ is the average vortex velocity will be induced
and a non-zero resistance will be measured. In the PE, some or all (reentrant
superconducting phase) the vortices are pinned again, resulting in a decrease
of the resistance or an increase of the critical current. The origin of the PE
is still under debate. Early, it was suggested to arise due to the softening of
the elastic moduli of the vortex lattice \cite{Pippard} and to a decrease of
correlation volume $V_{c}$ in the collective pinning theory of Larkin and
Ovchinnikov \cite{LO}. It has also been proposed to be the signature of a
disorder-induced or a thermally-induced order-disorder transition
\cite{Mikitik, PaltielPRL85, Beek, Park, Kleinnature413}. However, it has
equally been said to occur naturally at the crossover between a weak to strong
vortex pinning regime \cite{Blatter}. It could also simply appear depending on
the strength and density of pinning centers, and their competing action depending on magnetic
field and temperature \cite{XuPRL101, Werner, Kwok}. In general, the dependence of the PE on disorder is strongly dependent on the pinning mechanism: single vortex pinning, or collective.

In this work, we investigate how the PE depends on pinning strength by using a series of
Fe$_{x}$Ni$_{1-x}$Zr$_{2}$ metallic glasses with $x$ from 0 to 0.6, since changing $x$ modifies the pinning properties. The extreme purity and absence of long range order due to the amorphous nature of
these glasses confers them extremely weak pinning properties ($J_{c} \leq 0.4$ A/cm$^{2}$) which make
it an ideal system to study vortex phases and vortex motion. Pinning in these glasses is collective. In
Ref. \cite{HilkePRL91}, a huge PE, larger than in other weakly-pinned amorphous
systems \cite{BerlincourtPRL6, KesPRB28, GeersPRB63, WordenweberPRB33}, was
observed in a sample of Fe$_{0.3}$Ni$_{0.7}$Zr$_{2}$. Since the critical
current density in these alloys is at least ten times smaller than seen in other
amorphous alloys \cite{BerlincourtPRL6, KesPRB28, GeersPRB63,
WordenweberPRB33}, this provides an ideal case study in the weak-pinning limit.

The Fe$_{x}$Ni$_{1-x}$Zr$_{2}$ superconducting glasses are obtained by
melt-spinning, as described in Ref. \cite{DikeakosJNCS250}. Resistance
measurements are performed in the standard four-probe technique through
soldered indium contacts using a resistance bridge providing ac current at
15.9 Hz in a $^{3}$He refrigerator and a dilution refrigerator.

\begin{figure}
[ptbh]
\begin{center}
\includegraphics[
trim=0.000000in 4.697613in 0.000000in 0.000000in,
height=2.009in,
width=2.738in]%
{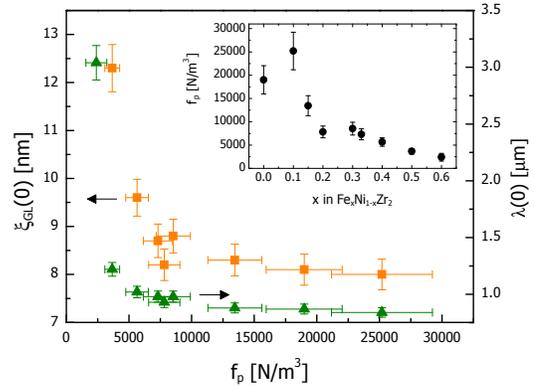}%
\caption{GL coherence length $\xi_{GL}\left(  0\right)  $ and penetration
depth $\lambda\left(  0\right)  $ as a function of the pinning force density
$f_{p}$. Inset: $f_{p}$ as a function of Fe content $x$ in Fe$_{x}$Ni$_{1-x}$Zr$_{2}$}%
\label{Fp}%
\end{center}
\end{figure}

\begin{figure}
[ptbh]
\begin{center}
\includegraphics[
trim=0.000000in 0.253381in 3.303984in 0.000000in,
height=6.3365in,
width=2.9386in
]%
{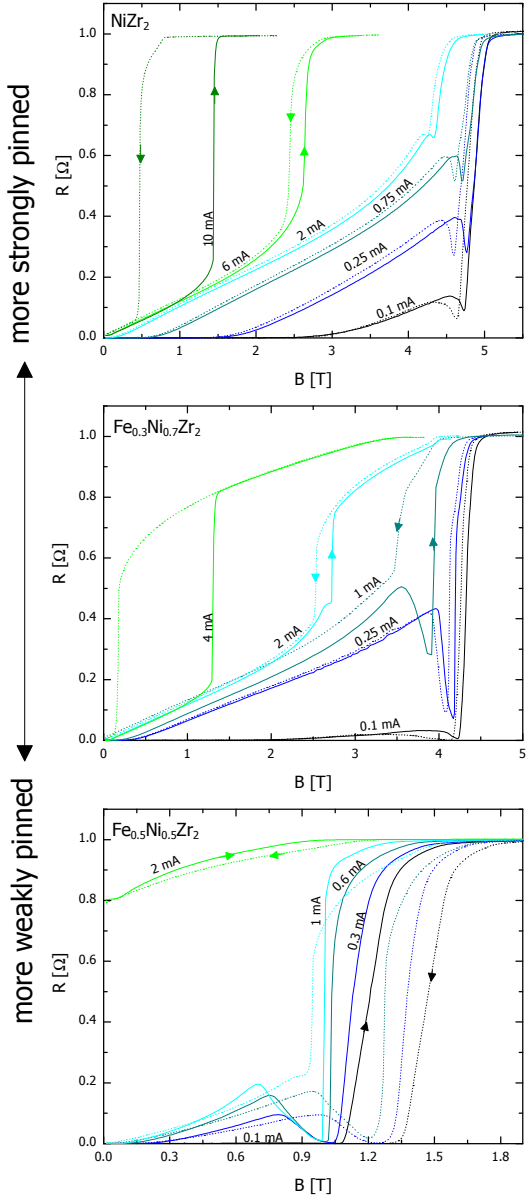}%
\caption{Resistance vs magnetic field for different driving currents as shown
at T$\approx$0.33 K measured on different metallic glasses of varying pinning
properties. a) NiZr$_{2}$ (more strongly pinned) b) Fe$_{0.3}$Ni$_{0.7}%
$Zr$_{2}$ c) Fe$_{0.5}$Ni$_{0.5}$Zr$_{2}$ (more weakly pinned).}%
\label{PE}%
\end{center}
\end{figure}

\begin{figure}
[ptbh]
\begin{center}
\includegraphics[
trim=0.000000in 6.126722in 0.382099in 0.000000in,
height=1.5454in,
width=2.6377in]%
{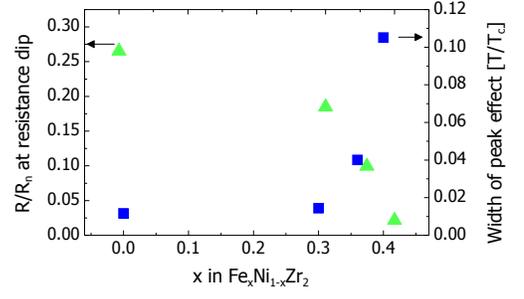}%
\caption{Lowest resistance reached in the resistance dip (PE) (left axis) and
width of the peak effect (right axis) as measured in temperature sweeps in a
fixed B field of 2 T with $I=0.1$ mA.}%
\label{PEinT}%
\end{center}
\end{figure}

The pinning force density is determined from $f_{p}=J_{c}B$, and is shown in the inset of Fig.\ \ref{Fp} as a function of Fe content $x$ at a magnetic field $B=0.15B_{c2}$, where $B_{c2}$ is the upper critical field. The critical current density $J_{c}$ is defined when the resistance reaches 0.5 m$\Omega$ which is close to our
experimental resolution. As can be seen, increasing the Fe content in these
glasses results in an important decrease of the pinning force density; $f_{p}$
is about a factor of five smaller in $x=0.5$ and $0.6$ than it is in $x=0.1$. This dependence can be related to the fact that with increasing $x$, vortex core and size increase substantially in these alloys. As calculated from expressions for superconductors in the dirty limit \cite{KesPRB28}, in $x=0$, $\xi_{GL}\left(  0\right) =8.1$ nm and $\lambda\left(  0\right) =0.87~\mu$m, which respectively almost doubles and triples in $x=0.6$ \cite{xdep}. This yields the relationship between $\xi_{GL}\left(  0\right)  $ and $\lambda\left(  0\right)  $, and $f_{p}$, shown in Fig.\ \ref{Fp}. A strong decrease of the pinning force density with increasing coherence length was also predicted by Larkin and Ovchinnikov \cite{LO} (LO) for 3D collective pinning, with dependence:
\begin{equation}
f_{p}=\frac{n^{2}\left\langle f^{2}\right\rangle ^{2}}{10B^{2}C_{66}^{2}%
\xi^{3}}\text{,}\label{fpxi}%
\end{equation}
where $n$ and $f$ are the density and strength of pins respectively, and $C_{66}$ is the shear modulus which describes the elasticity of the vortices. The decrease of $f_{p}$ with increasing $ \xi$ is readily understood considering that, for identical vortex number density, vortex overlap is enhanced for large $ \xi$ and $\lambda$, thereby increasing collective interactions between vortices which tend to
order the system and reduce pinning. Indeed, according to the LO collective pinning theory \cite{LO}, the scale of vortex interactions can be described by a correlation volume $V_{c}=R_{c}^{2}L_{c}$, where $R_{c}$ and $L_{c}$ are the correlation radius and length respectively. $R_{c}$ increases with $ \xi$ according to $R_{c}=4\pi^{1/2}BC_{66}^{3/2}\xi^{2}/n\left\langle f^{2}\right\rangle $, which then results in a decrease of the pinning force
\begin{equation}
f_{p}=\left(  \frac{n\left\langle f^{2}\right\rangle }{V_{c}}\right)
^{1/2}\text{.}\label{fpVc}%
\end{equation}
According to the LO collective pinning theory, in the weak pinning limit and for high magnetic fields where the number of vortices greatly surpasses the number of defects, reentrant pinning in the peak effect results from the pinning of mobile vortices through collective interactions with pinned vortices. Hence, enhanced collective interactions strengthen the peak effect.

\begin{figure}
[ptb]
\begin{center}
\includegraphics[
height=4.2194in,
width=3.2214in
]%
{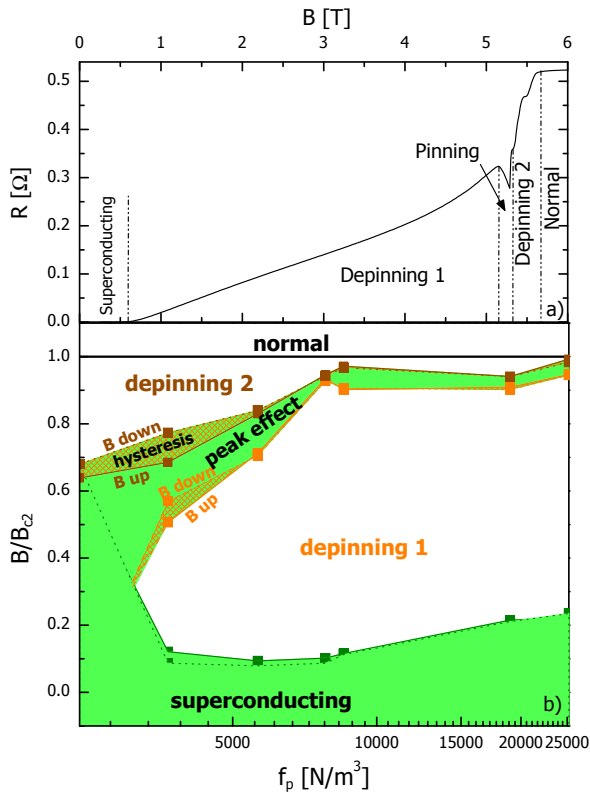}%
\caption{a) R vs B trace measured at T=0.35 K on Fe$_{0.1}$Ni$_{0.9}$Zr$_{2}$
with $I=1$ mA showing how different vortex phases are defined. b) Phase diagram
of vortex dynamics as a function of pinning force density. The phase
boundaries are defined as described in the text and extracted from R vs B data
measured with J$\approx$1.8 A/cm$^{2}$ for each alloy composition. The solid
lines represent phase boundaries obtained in increasing B field sweeps while
the dotted lines are for decreasing field sweeps. In this manner, we can
identify regions of hysteresis (hatched areas).}%
\label{phase}%
\end{center}
\end{figure}

The evolution of pinning strength and collective vortex interactions with Fe
content in Fe$_{x}$Ni$_{1-x}$Zr$_{2}$ evidenced above allows us to study how the peak effect depends on collective effects. As seen in resistance as a function of magnetic
field measurements on samples with $x=0,~0.3$ and $0.5$ in Fig.\ \ref{PE}, the
dip in resistance characterizing this B-induced re-entrant pinning phase
changes from thin and shallow in the more strongly pinned NiZr$_{2}$ to very broad and deep in the more weakly pinned 
Fe$_{0.5}$Ni$_{0.5}$Zr$_{2}$. In the later case, the resistance even decreases
back to zero in the peak effect and a large re-entrant
superconducting phase is seen. In crystals of 2H-NbSe$_{2}$ containing a large amount of quenched
disorder introduced as Fe impurities and showing stronger point pinning properties, a broadening of
the PE has been observed before \cite{BanerjeePRB59}. This is mainly due to an increased inhomogeneous pinning where the preferred pinning of single vortices impedes the others. This can be contrasted to our amorphous alloys in which pinning is weak but collective and PE broadening is seen with decreasing pinning force and growing collective effects.

A similar broadening of the PE with $x$, with accompanied deepening of the resistance dip at the PE is also observed in resistance measurements performed as a function of temperature in a fixed B field (Fig.\ \ref{PEinT}). This important observation confirms that PE broadening is indeed attributable to the augmentation of collective interactions and not due to the presence of inhomogeneities in the material. This confirmation is essential here because in Ref.\ \cite{xdep}, we have evidenced the presence of structural inhomogeneities, especially in alloys with a relatively large Fe content $x=0.5$ and $x=0.6$, which causes an inhomogeneous distribution of vortices upon entry and exit in a B field sweep. This effect is dynamic; in a constant magnetic field, the vortex distribution eventually becomes homogeneous as the vortices diffuse throughout the more weakly and strongly pinned regions of the material. As a result, during a temperature sweep in a constant B field, the flux distribution remains homogeneous as the number of vortices in the sample is fixed. It was suggested by Brandt \cite{Brandt} that a PE could arise in superconductors in which large inhomogeneities, with $T_{c}$ higher than that of the main phase, are present. An enhancement of the PE in simulated systems of weakly-pinned interacting polydisperse Yukawa particles was also observed by Reichhardt and Reichhardt \cite{ReichhardtPRE}.

Having established that the characteristics of the PE as a function of $x$ are equivalently observed in B field and temperature sweeps, we complete our analysis from measurements performed as a function of B field because these results are more readily obtained. A typical $R$ vs $B$ trace is shown in Fig.\ \ref{phase}a). Following the
naming scheme of Ref. \cite{HilkePRL91} to identify vortex phases, we distinguish the superconducting phase where $R=0$ at low $B$ field, followed at higher field by a
depinned vortex phase called depinning 1, the onset of which is defined when
the resistance reaches 0.5 m$\Omega$. In Refs. \cite{LefebvrePRB74,
LefebvrePRB78}, we have demonstrated that the depinning 1 phase is
characterized by the long range ordered moving Bragg glass phase (MBG)
\cite{GiamarchiPRB57}. At still higher $B$ field, a reentrant pinning phase is
seen; the onset of this phase is defined when $dR/dB=0$, and its termination is defined
at the position in $B$ where the resistance reaches the same value as at the onset of
the pinning phase. The end of the pinning phase marks the onset of the
depinning 2 phase, which we have shown in Ref. \cite{LefebvrePRB74} exists
even in the lowest driving current regime. This phase has smectic order
characteristic of the moving transverse glass (MTG) \cite{GiamarchiPRB57} in
which the orientation of channels in which vortices flow can vary suddenly
depending on the driving force and vortex density. Finally B$_{c2}$ is defined
at the point of strongest negative curvature before reaching the normal state.

Extracting the boundaries of vortex phases according
to the definitions above from $R$ vs $B$ traces measured at
constant current density J$\approx$1.8 A/cm$^{2}$, we obtain the phase diagram of Fig.\ \ref{phase}b). For
all alloys, this current density corresponds to a regime in which we observe a
peak effect, and never a direct transition from the depinning 1 to the
depinning 2 phase as visible for example in the $I=10$ mA trace in Fig.\ \ref{PE}a). In the diagram, the superconducting and peak effect phases are
represented by filled green areas. These two phases merge in the most weakly pinned sample ($x=0.6$) in
which no PE is visible and the vortices remain pinned up to the depinning 2
phase. A downward bending of the pinning phase toward lower reduced field b=B/B$_{c2}$ with decreasing pinning force is observed. In the PE, an amorphization of the vortex lattice with collapse of $V_{c}$ occurs \cite{LO} due to a softening of the elastic moduli \cite{Pippard}, which increases pinning. In the high $B$ field range where the PE appears, the size of the correlation radius $R_{c}$ becomes comparable to $\xi$ and to the inter-vortex distance $a$. In the large coherence length limit, where $R_{c}$ is also largest, the onset of the peak effect can occur at lower magnetic field where $a$ is larger, which explains this downward bending of the pinning phase toward lower $b$ in this limit. In the most weakly pinned sample, $V_{c}$ presumably becomes so large due to the large $\xi$ and $\lambda$ that, even at very low $B$, coherent pinning of these large vortex bundles does not permit depinning. As a result, the sample remains in the pinning phase up to the transition to the depinning 2 phase. The broadening of the PE phase with decreasing $f_{p}$ (or increasing $V_{c}$) is readily seen from the phase diagram and confirms that reentrant pinning is strengthened by collective vortex interactions. This also infers that collective vortex interactions cause the PE in these materials.

In the phase diagram of Fig.\ \ref{phase}b), the solid and dotted lines represent transitions obtained in increasing and decreasing magnetic field sweeps respectively. As a result, regions of hysteresis become visible, as highlighted by the orange hatched areas in the weak pinning range. These hysteresis regions are not stable and depend on the $B$ field sweep rate. They arise due to the inhomogeneous distribution of vortices resulting from
structural inhomogeneities discussed earlier. However, even ignoring the hysteresis regions, the broadening of the PE with decreasing pinning force is obvious. 
A large broadening of the depinning 2 phase is also visible in the low pinning force region of the phase diagram of Fig.\ \ref{phase}b). This broadening is partly due to the increasingly two-phase character of these alloys. An increase of the $B_{c2}$ transition width is common in inhomogeneous superconductors. At this stage, it is not known how the intrinsic pinning force and collective vortex interactions affect the smectic order characteristic of the depinning 2 phase and if it could cause its widening.

In summary, we have presented the pinning force dependence of the peak effect based on measurements in
the metallic glasses Fe$_{x}$Ni$_{1-x}$Zr$_{2}$. It was shown that in this metallic glass series, the intrinsic pinning force decreases with Fe content as the coherence length and penetration depth increase, as well as collective vortex interactions. Then, a strengthening of the peak effect, which broadens to eventually fill the whole space below the transition to the depinning 2 phase in the most weakly pinned sample, was seen with decreasing pinning force. These observations confirm that collective vortex interactions are at the origin of the peak effect phenomenon in these weakly-pinned metallic glasses.

\end{document}